\documentclass[%
 reprint,
 amsmath,amssymb,
prl,
]{revtex4-1}

\usepackage{graphicx}
\usepackage{dcolumn}
\usepackage{bm}
\usepackage{epstopdf}
\usepackage{color}
\usepackage{amsmath}

\begin{document}
\newcommand{\todo}[2]{\textcolor{red}{\textsc{#1}}}
\preprint{APS/123-QED}
\title{A solid state single photon source with Fourier 
Transform limited lines at room temperature} 

\author
{A. Dietrich$^1$, M.W. Doherty$^2$, 
I. Aharonovich$^3$, A. Kubanek$^{*,1,4}$\\}
\affiliation{$^1$ Insitute for Quantum Optics, Ulm University,D-89081 Ulm, Germany\\
 $^2$Laser Physics Centre, Research School of Physics and Engineering, Australian National University, Australian Capital Territory 2601, Australia,\\
 $^{3}$School of Mathematical and Physical Sciences, Faculty of Science, University of Technology Sydney, Ultimo, New South Wales 2007, Australia\\
$^4$ Center for Integrated Quantum Science and Technology (IQst),Ulm University, D-89081 Ulm, Germany\\
$^\ast$To whom correspondence should be addressed; E-mail:  alexander.kubanek@uni-ulm.de
 }

\date{\today}

\begin{abstract}

Solid state single photon sources with Fourier Transform (FT) limited  lines are among the most crucial constituents of photonic quantum technologies and have been accordingly the focus of intensive research over the last several decades.
However, so far, solid state systems have only exhibited FT limited lines at cryogenic temperatures due to strong interactions with the thermal bath of lattice phonons.
In this work, we report a solid state source that exhibits FT limited lines measured in photo luminescence excitation (sub 100 MHz linewidths) from 3K--300K.
The studied source is a color center in the two-dimensional hexagonal boron nitride and we propose that the center's decoupling from phonons is a fundamental consequence of material's low dimensionality.
While the center's luminescence lines exhibit spectral diffusion, we identify the likely source of the diffusion and propose to mitigate it via dynamic spectral tuning. 
The discovery of FT-limited lines at room temperature, which once the spectral diffusion is controlled, will also yield FT-limited emission.
Our work motivates a significant advance towards room temperature photonic quantum technologies and a new research direction in the remarkable fundamental properties of two-dimensional materials.
\\
\begin{description}

\item[DOI] XXXXXXXX   

\end{description}
\end{abstract}

\maketitle


Future applications of quantum technology rely on compact and scalable quantum platforms.
An essential building block for photonic quantum technologies, including quantum communication networks, quantum sensors and distributed quantum networks, are single photon emitters (SPEs)\cite{atature_material_2018,awschalom_quantum_2018,lodahl_interfacing_2015,gao_coherent_2015}.
Coherent photon absorption and emission is a fundamental prerequisite to ensure, for example, coherent read and write of quantum information in  various quantum protocols.
Perfect coherence is achieved when all incoherent processes arising from interactions with the environment are suppressed.
Once suppressed, the spectral line of an SPE matches the Fourier Transform of its excited state decay.
\\
This so-called Fourier Transform (FT) limit can be achieved at present with cold atomic ensembles \cite{qian_temporal_2016}, Doppler-broadened atomic ensembles \cite{jeong_quantum_2017}  and single cold atoms \cite{kuhn_deterministic_2002}, enabling single photon absorption and generation with high efficiency.
However, cold atoms are limited in their photon absorption and generation rate and require complex apparatus for atom trapping and cooling in ultra-high vacuum conditions. All of which is challenging to scale.
To achieve scaling via a different approach, various solid state quantum systems have been thoroughly investigated as SPEs.
To date, these systems are constrained to cryogenic conditions where interactions with the solid-state environment, in particular the thermal bath of lattice phonons, are adequately suppressed.
Even at very low temperatures, only a few systems, such as III-V  quantum dots (QDs) \cite{sapienza_nanoscale_2015,wang_-demand_2019,senellart_high-performance_2017}, color centers in solids \cite{sipahigil_quantum_2012,bhaskar_quantum_2017,tran_quantum_2016} as well as single molecules \cite{chu_single_2017}, have shown FT limited spectral lines.
\\
Here we report a solid-state SPE in a two-dimensional (2D) material -- hexagonal boron nitride (hBN) -- that exhibits FT limited lines at room temperature under resonant excitation (see figure 1(d)).
This exceptional observation constitutes a significant advance in solid state SPEs and implies that this SPE is decoupled from the low-frequency phonons that broaden other solid state SPEs.
We propose that this decoupling occurs as a consequence of an out-of-plane distortion of the SPE, such that it no longer strongly interacts with the in-plane phonon modes.
All in-plane color centers in 2D materials have the potential to undergo such a distortion due to the intrinsic instability of an in-plane configuration and the variety of ways that planar symmetry can be locally removed (e.g. via nearby defects, boundaries and interfaces) to bias a particular out-of-plane direction (e.g. \cite{noh_stark_2018}).
Consequently, it is possible that decoupling of color centers from phonons is a fundamental feature of 2D materials and thus our observations motivate a new research direction across color centers in 2D materials.
\\
A variety of SPEs have been recently discovered in hBN \cite{tran_quantum_2016,jungwirth_temperature_2016,exarhos_magnetic-field-dependent-2019,proscia_near-deterministic_2018,konthasinghe-rabi-2019}.
These SPEs feature ultra bright and polarized emission with extremely high Debye-Waller factors.
Some SPEs have even shown a magneto-optical effect consistent with an optically-addressable metastable electron spin, which is a valuable resource in quantum information processing \cite{exarhos_magnetic-field-dependent-2019}.
There are also a variety of benefits of the 2D nature of the material, including efficient optical collection and flexible device design, including the diverse opportunities offered by heterostructures of van der Waals materials.
Indeed, the layered nature of the material has already been exploited to realize spectral tuning of the ZPLs of the SPEs, using mechanical induced strain \cite{grosso_tunable_2017} or Stark tuning \cite{noh_stark_2018}, over a range of 1.4 THz.
However, owing to their variety and, until recently, issues with their reproduction, the chemical structure of the SPEs in hBN is unknown.
\\
The SPE we studied is a standard defect in hBN produced by thermal annealing and was found in a multi-layer hBN flake (hight $\approx$ 100--300 nm). 
It is characterised by a photoluminescence (PL) zero-phonon line (ZPL) at 635.5 nm that has a full-width half-maximum (FWHM) linewidth of 177 GHz and a Debye-Waller factor of 0.8 under far off-resonance excitation (532 nm) at 3 K (see figure 1(b)).
As demonstrated in figure 1(b), the PL phonon sideband (PSB) has a feature at 48 THz, which we address to optical phonon mode \cite{dietrich_observation_2018,tran_resonant_2018}.
\begin{figure}[htbp]
\centering
\includegraphics[width=0.48\textwidth]{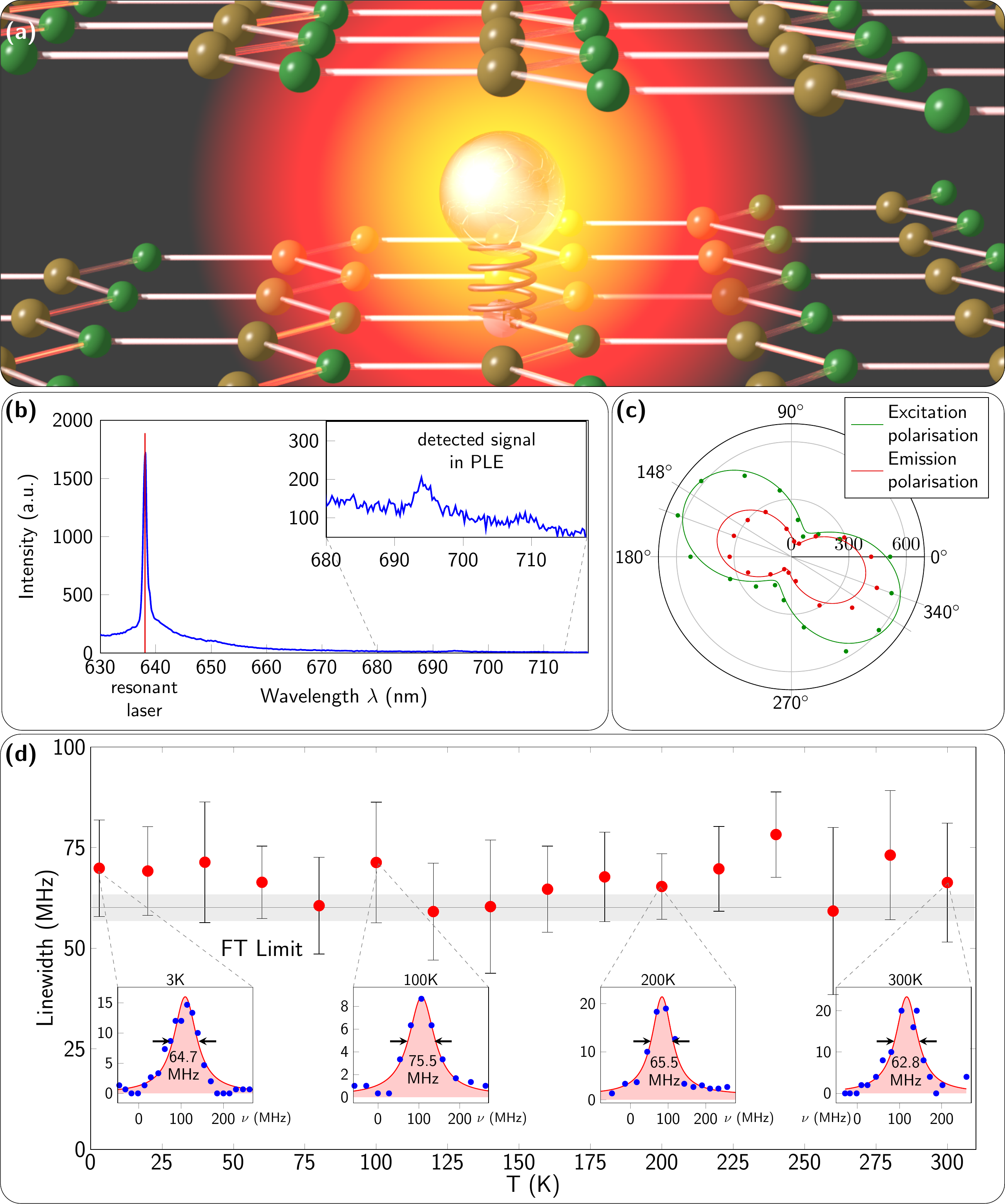}
{\centering}
\caption{\label{fig:fig1}\textbf{(a)} Atomic defect in hexagonal boron nitride that acts as a single photon emitter. The electronic wavefunction of the defect center embedded in the hexagonal structure of hexagonal Boron Nitride is illustrated as a ball-and-stick model. \textbf{(b)} The PL spectrum of the emitter. Inset: the PSB features that were used for the detection of the PLE signal via their selection using a 680 nm long-pass filter. \textbf{(c)} The polarization-dependence of the dipole orientation shows an offset of 12$^\circ$ when the polarize is placed in the off-resonant excitation path or in the emission path (details see \cite{dietrich_observation_2018}). \textbf{(d)} The homogeneous linewidth in PLE over temperature. The linewidth stays close to the FT limit of $60.1 \pm 3.4$ MHz in a temperature range from 3 K up to room temperature. The recorded PLE spectra with the Lorentzian fits are depicted for 3 K, 100 K, 200 K and 300 K.}
\end{figure}
\\
In the following, we first present the results and analysis of three different experiments on the SPE: photoluminescence-excitation (PLE) spectroscopy, resonant excitation photodynamics, and PL spectroscopy; each conducted over the temperature range 3--300 K.
We then propose a model of the SPE that seeks to explain its decoupling from low-frequency acoustic phonons as well as a model of its spectral diffusion mechanism and its dependence on temperature, excitation power and wavelength.
\\
Our PLE spectroscopy was performed by scanning the frequency of a laser through resonance with the SPE and recording the SPE's PSB fluorescence as a function of time, as spectrally isolated with the aid of a long-pass filter (see figure 1 and SM for experimental details).
For scans of duration $<0.03$ s, the observed ZPL has a Lorentzian lineshape at all temperatures, but the precise position of the line differs between each scan due to spectral diffusion.
The consistent Lorentzian shape of the lines demonstrates that the scans are sufficiently fast to capture the true homogeneous lineshape and not the consequence of spectral jumps occurring during scans, which would not yield such consistently symmetrical lineshapes (see SM for a detailed analysis of the lineshape variation with scan speed).
Accepting the observed lines as the true homogeneously broadened lineshape of the SPE, we extract the FWHM linewidths and find that, across the full temperature range up to 300 K, they are approximately constant, well below 100 MHz, and within measurement uncertainty of the FT limit 60.1$\pm$ 3.4 MHz.
The latter being predicted by the previously observed excited state lifetime of 2.65 ns \cite{dietrich_observation_2018}.
\\
The FT limited width of the PLE ZPL implies that there is no coupling to low frequency acoustic phonons (or any other dephasing mechanism) on the timescale of the laser scans.
Additionally, the temperature independence of the FT limited linewidth is consistent with previous observations \cite{jungwirth_temperature_2016,kianinia_robust_2017} that lifetimes of SPEs in hBN are temperature independent.
We indirectly confirmed this to also be the case for this SPE via second-order correlation measurements (see SM).
\\
\\
\begin{figure*}
\centering
\includegraphics[width=\textwidth]{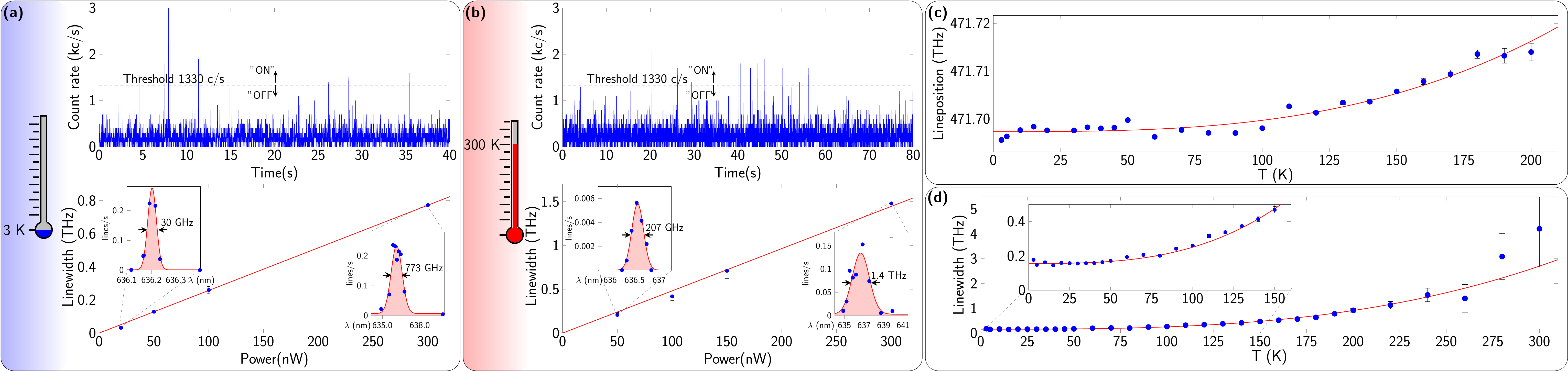}
\caption{\label{fig:fig2}\textbf{(a)} Fluorescence timetrace under resonant excitation (upper panel) and power-dependency of the inhomogeneous linewidth (lower panel) at 3 K. \textbf{(b)} Fluorescence timetrace under resonant excitation (upper panel) and power-dependency of the inhomogeneous linewidth (lower panel) at 300 K. A threshold (depicted in the fluorescence traces) is defined to distinguish between signal and noise and to estimate the amount of spectral diffusion that is the origin for inhomogeneous line broadening (see SM  for details). The inhomogeneous linewidth shows a linear power dependency with a linewidth as broad as 773 GHz at 3K and 300 nW excitation power and 1.4 THz at 300 K. \textbf{(c)} Temperature-dependence of the PL ZPL energy. \textbf{(d)} Temperature-dependence of the PL ZPL linewidth. Solid lines in \textbf{(c)} and \textbf{(d)} are fits of the function $a+bT^3$, where $a=471.6973(3)$ THz and $b=2.4(1)\cdot10^{-9}$ $\mathrm{THz/T^3}$ for the shift and $a=156(6)$ GHz and $b=9.4(2)\cdot10^{-5}$ $\mathrm{GHz/T^3}$ for the width.}
\end{figure*}
It is natural to ask how the absence of coupling to acoustic phonons is consistent with the presence of the coupling to high frequency optical phonons that yields the PSB features, which we used for PLE detection.
There are two key reasons.
Firstly, the optical phonons will not yield observable affects on the ZPL until very high temperatures, owing to their high frequencies.
Secondly, the optical phonons can couple to the defect at a distance due to the electric fields generated by their dynamic polarization.
Thus, they are not limited to short-range deformation potential interactions with the SPE like acoustic phonons and so are able to interact with the SPE even when the acoustic phonons are decoupled.
\\
To investigate the spectral diffusion giving rise to the jumps of the ZPL between the laser scans of the PLE spectroscopy, we performed resonant excitation photodynamics measurements at 3 and 300 K by fixing the laser frequency near resonance and measuring the PSB fluorescence as a function of time in a similar fashion as before.
Owing to the spectral diffusion, we see fluctuations in the detected fluorescence with time as the SPE's transition randomly jumps in and out of resonance with the laser (see figures 2(a) and (b)).
For each time-trace, we apply a signal threshold conditioned on the signal level being six times above the noise level (see SM for discussion).
We find that the inhomogeneous distributions are best described as Gaussian with FWHM linewidths that depend linearly on excitation power, as depicted in figures 2(a) and (b) (lower panels).
The linear gradient of the linewidth with power depends slightly on temperature (less than a factor of 2 from 3 -- 300 K), by increasing the temperature the gradient increases correspondingly.
By extrapolating to zero excitation power, the inhomogeneous linewidth approaches the FT limit at both 3 and 300 K, admittedly with large error margins of 10 and 50 GHz, respectively.
\\
To further investigate the spectral diffusion, we performed PL spectroscopy under far off-resonance excitation (532 nm, 30 $\mu$W) across the temperature range 3--300 K.
We found the PL ZPL shapes to be Gaussian and fitted them to extract the line position and width as a function of temperature.
The ZPL width varies from 177 GHz at 3 K up to $\approx$ 4 THz at 300 K.
The ZPL position similarly shows a significant shift from 471.65 THz at 3 K to 471.83 THz at 200 K.
As demonstrated by the fits in figures 2(c) and (d), both the width and position are consistent with a $T^3$ temperature dependence up to near room temperature.
At both 3 K and 300 K, the PL linewidths are comparable to the inhomogeneous linewidths observed under resonant excitation.
This provides an unambigios proof that the observed signal corresponds to the studied SPE.
\\
Each of these observations support the conclusion that the PL ZPL shape arises from spectral diffusion and not from an onset of electron-phonon interactions under the far off-resonance excitation.
There are several key features of the broadening and shift that point away from electron-phonon interactions. The first is that the shift is towards higher energy with increasing temperature, which differs from previous observations of other SPE in hBN \cite{jungwirth_temperature_2016}.
This would be highly unusual for electron-phonon interactions, as it would imply that the phonons have higher frequencies in the excited electronic state than the ground electronic state, in contrary to the usual case where the frequencies are higher in the ground state as lower energy electronic wavefunctions are typically more spatially confined, leading to stiffer bonds.
Secondly, it is highly unusual for the broadening and shift to have similar temperature power series since different electron-phonon interactions and processes give rise to spectral shift (i.e. first-order quadratic interactions) and broadening (i.e. Raman-type processes of linear interactions \cite{davies_jahn-teller_1981,stoneham_theory_2001}).
Rather, as will be discussed, shifts in line positions to higher energy and correlations in line widths and positions is common in inhomogeneous broadening mechanisms.
\\
To explain the SPE's decoupling from low-frequency acoustic phonons, we propose that it is an in-plane defect that has undergone an out-of-plane distortion to a configuration where the electron orbitals (i.e. the highest occupied molecular orbital (HOMO) and lowest unoccupied molecular orbital (LUMO)) involved in the emitters optical transition no longer interact strongly with the in-plane acoustic phonons.
This is inspired by the calculations recently reported by Noh et al. \cite{noh_stark_2018} that showed that defects of the class $\mathrm{X_BV_N}$ (where $\mathrm{X_B}$ is a substitutional impurity at a boron site and $V_N$ is a vacancy at a nitrogen site) possess a double-well vibrational potential energy function for the out-of-plane displacement of the defect.
The two wells corresponding to displacements below and above the plane, with a local maximum at the in-plane position (figure 3).
This may be understood as the out-of-plane displacement providing an additional degree of freedom that the defect can exploit to minimise its total energy by elongating and reorientating its bonds.
The latter constituting a shift from pure sp$^2$ bonding to sp$^3$ or other bond types.
If the reflection symmetry of the plane is maintained, then these two wells will be identical and the vibronic states of the defect will be superpositions of the vibrational wavefunctions of the two wells (analogous to the well-known inversion doubling of ammonia), such that the defect's mean position dynamically remains in-plane.
However, if reflection symmetry of the plane is locally removed, by say interactions with a nearby surface or defect, then one of wells will lower in energy and shift further away from the plane, while the other will do the opposite, thereby shifting the defect's dynamic mean position out of plane (see \cite{noh_stark_2018} for explanation in context of the Stark effect).
If the reduction in the well's energy is sufficiently large, this will result in a static out-of-plane distortion.
\begin{figure}[htbp]
\centering
\includegraphics[width=0.48\textwidth]{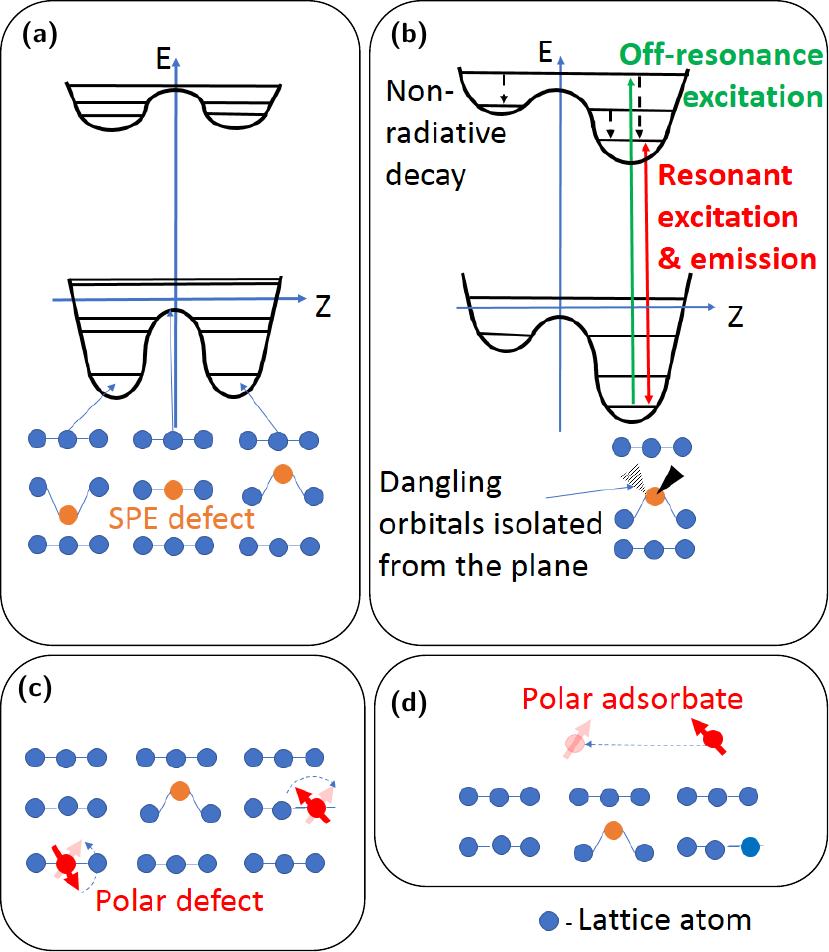}
\caption{\label{fig:fig3}{\bf (a)} Configuration coordinate diagram depicting the double-well vibrational potential energy of the ground and excited electronic states as a function of the defect's out-of-plane displacement. {\bf (b)} A similar diagram for the case where reflection symmetry has been lowered such that the defect statically distorts out of plane. Processes of optical ZPL emission and resonant and far off-resonance excitation are depicted, including indicative transition dipole moments. {\bf (c)} and {\bf (d)} Sketches of the spectral diffusion sources of defects within the hBN changing their polarization upon optical excitation and polar molecules hopping between adsorbate sites upon optical excitation.}
\end{figure}
\\
Decoupling from in-plane acoustic phonons in this distorted configuration will occur if the HOMO and LUMO are predominately formed from defect atomic orbitals that are not participating in covalent bonding to the plane (see figure 3).
For example, the dangling sp$^3$ orbitals resulting from the shift away from sp$^2$ bonding due to the distortion or the low energy $d$ orbitals of an impurity etc.
Thus, relative displacement of the atoms associated with in-plane acoustic modes will not significantly affect the HOMO and LUMO, and so they may be pictured as isolated orbitals, like those of a trapped atom.
Notably, the existence of SPE HOMOs and LUMOs with orbitals not participating in covalent bonding is permitted by the separation of hBN layers, which is sufficiently large to encompass an atomic orbital, and the absence of covalent bonding of the layers.
\\
This out-of-plane distortion model also offers an explanation of another puzzling observation -- that the emission polarization is different to the far off-resonance excitation polarization (see figure 1(c)) -- which has also been made for other SPEs in hBN, but is yet to be explained.
Figure 3 depicts the optical transitions that correspond to emission, resonant and far off-resonance excitation.
Emission and resonant excitation occurs between vibronic states associated with the lower energy potential well, whereas the final vibronic state of far off-resonance excitation may extend into the higher energy potential well.
The two sets of transitions will thus have different dipole moments due to the differences in the defect configurations involved.
\\
Since each of the above arguments can be similarly applied to SPEs in other 2D materials, we believe that decoupling of SPEs from low-frequency phonons is a fundamental feature of 2D materials.
The key ingredients being the layered planar geometry of 2D materials, absence of covalent bonding between layers, and SPEs that undergo out-of-plane distortions, where there HOMOs and LUMOs that do not participate in bonding with the plane.
Future simulations and experiments could test this model by calculating or measuring the phonon coupling of a SPE as its out-of-plane distortion is manipulated via electric fields or compression.
\\
Turning now to an explanation of the SPE's spectral diffusion, we begin with the hypothesis that the spectral jumps of our SPE may  arise from fluctuations in the local electric field due to nearby defects undergoing random transitions betweens states with different charges or polarizations under optical excitation.
Alternatively, the fluctuations could arise from charged or polar molecules hopping between adsorption sites on the hBN surface under optical excitation, which can be equivalently pictured as a site undergoing random transitions between occupied and unoccupied states.
If these transitions are also temperature and wavelength dependent, then the distribution of SPE's spectral jumps would likewise be dependent on excitation power, wavelength and temperature, as we have observed.
\\
In the SM, we apply statistical models established in the literature for 3D solids \cite{stoneham_shapes_1969,davies_approximate_1971}. These models  show that only defects and polar adsorbate changing their polarization and hopping is consistent with the observed power and temperature dependence. 
However, we note that these models do not fully capture the anisotropy of hBN and potential non-uniform distribution of defects and adsorption sites in the vicinity of the SPE.
\\
This conclusion can be tested in the future by observing the change in spectral diffusion with changes in dielectric environment (i.e. by placing the hBN onto substrates with different permittivities) or atmosphere (i.e. varying humidity). 
Furthermore, this conclusion implies that these sources of spectral diffusion can be reduced by eliminating defects through refined hBN fabrication and encapsulating the hBN within dielectric structures designed to reduce quasi-static electric fields within the hBN. 
After sufficient reduction, the SPE line can be dynamically stabilized via Stark tuning \cite{noh_stark_2018,acosta_dynamic_2012,de_las_casas_stark_2017} and so truly realize a stable FT limited single photon source at room temperature for large-scale photonic quantum technologies. 
\\
Our results constitute major advance in the field of quantum photonics and open up new opportunities for solid state platforms at room temperature.
First, single atom-photon quantum interfaces could be potentially achieved with a solid state-state platform at room temperature facilitating novel approaches towards quantum networks or quantum sensing. 
Second, it stimulates the community to further investigate the level structure of quantum emitters in hBN to understand the underlying mechanisms and to enable the deterministic engineering of other systems with similar behaviour. 
In the context of novel quantum materials our platform enables to combine the attractive mechanical, electrical, thermal and chemical \cite{paszkowicz_lattice_2002} properties of two-dimensional materials with the atom-like optical properties of a quantum emitter at room temperature.
hBN is a key component in many van der Waals heterostructures, and can now be combined with ideal SPEs, e.g. for photon-mediated interactions.
Finally, the observed spectral diffusion can be further eliminated by surface termination, electric field stabilization \cite{nikolay_very_2019} or by embedding the  two-dimensional material into a protecting host matrix, ultimately leading to FT limited spectral lines in the emission process at room temperature.
Regardless, even with the current rate of emission, the hBN flakes can be interfaced with high collection antennas or cavities to progress with integrated solid state quantum photonics.
\\
\begin{acknowledgments}
AK acknowledges the generous support of the DFG, the Carl-Zeiss Foundation, IQST, the Wissenschaflter-R\"uckkehrprogramm GSO/CZS and the EFRE-Programm Baden-W\"urttemberg.
I.A. acknowledges the generous support of the Alexander van Humboldt Foundation, and the Asian Office of Aerospace Research \& Development (grant \# FA2386-17-1-4064), the Office of Naval Research Global (grant \# N62909-18-1-2025) and the Australian Research Council (DP180100077).
M.W.D. acknowledges support from the Australian Research Council (DE170100169). We thank Fedor Jelezko, Stefan H\"au\ss ler, Michael H\"ose, Rebecca Bernsdorff, Prithvi Reddy and Hosung Seo for fruitful discussions and experimental support.

\end{acknowledgments}

\appendix


\nocite{*}
\bibliography{A_solid_state_single_photon_source_with_FTL_at_RT_Arxive}

\end{document}